\documentclass[]{./raa_rb}
\usepackage{amssymb}
\usepackage{amsmath}
\usepackage{graphicx,times}
\input{psfig}

\begin{document}

\title{Grids of stellar models including second harmonic and colours: Solar composition
($Z=0.0172$, $X=0.7024$)}
   \volnopage{Vol.0 (200x) No.0, 000--000}      
   \setcounter{page}{1}          

\author{M. Y{\i}ld{\i}z}

\institute{Ege University, Department of Astronomy and Space Sciences, Bornova, 35100 \.Izmir, Turkey;
{\it mutlu.yildiz@ege.edu.tr}}

 \date{Received~~2009 month day; accepted~~2009~~month day}

\abstract{ 
Grids of stellar evolution are required in many fields of astronomy/astrophysics,
such as planet hosting stars, binaries, clusters, chemically peculiar stars, etc.
In this study, a grid of stellar evolution models with updated ingredients and
{recently determined solar abundaces} is presented. The solar values for the initial abundances of
hydrogen, heavy elements and mixing-length parameter are 0.0172, 0.7024 and 1.98,
respectively. The mass step is small enough (0.01 M$_\odot$)
that interpolation for a given star mass is not required.
The range of stellar mass is 0.74 to 10.00 M$_\odot$.
We present results in different forms of tables for easy and
general application. The second stellar harmonic, required for analysis of apsidal
motion of eclipsing binaries, is also listed. We also construct rotating models to
determine effect of rotation on stellar structure and
derive fitting formula for luminosity, radius and the second stellar harmonic as a
function of rotational parameter.  We also compute and list colours and bolometric
corrections of models required for transformation between theoretical and observational
results.  The results are tested for the Sun, the Hyades cluster, the slowly rotating
chemically peculiar Am stars and the eclipsing binaries with apsidal motion.
The theoretical and observational results along isochrones are in good agreement. The grids are also
applicable to rotating stars provided that equatorial velocity is given.
} 
\titlerunning{Grids of stellar models}
\authorrunning{M. Y{\i}ld{\i}z}

   \maketitle

\section{Introduction}


Mankind has often wondered what phenomena may lie 
behind the visible part of the universe. In this connection,
we observe the surface of stars and try to understand what happens inside them and what influences 
their structure. This is essential for our comprehensive description of the universe. 
Updated grids for stellar models are required for this task (e.g., Yi, Kim, \& Demarque 2003; Pietrinferni et al. 2006; VandenBerg et al. 2006;
Dotter et al. 2008). The present study 
focuses on constructing upgraded stellar models (rotating and non-rotating) 
with very dense mass steps.

The most important target for a stellar evolution code is the Sun,
due to the wealth of high quality seismic and
non-seismic constraints. The agreement between the Sun and the calibrated solar model
is a measure of code quality. 
{Such a solar model is obtained by using recently 
determined solar composition (Asplund et al. 2009)}. The relative sound speed difference between the Sun and the standard solar model constructed by using 
the {\small ANK\.I} (ANKARA-\.IZM\.IR; Ezer \& Cameron 1965) code  is less than 1.4 per cent (Y{\i}ld{\i}z 2008).
For a non-standard solar model constructed with enhanced opacity, however, the maximum relative sound speed 
difference is about 0.15 per cent (Y{\i}ld{\i}z 2011a).
The input parameters for construction of stellar interior are mixing length parameter ($\alpha$), 
initial hydrogen ($X$), and heavy element ($Z$) abundances. They are taken as the solar values: {$X=0.7024$, $Z=0.0172$, and $\alpha= 1.98$}. 
 
The present grids have some distinctive feature: 1) The code used to construct models of stellar interior is 
very successful in modeling the Sun, $\alpha$ Cen A and B
(Y{\i}ld{\i}z 2007; 2011a), eclipsing binaries (Y{\i}ld{\i}z 2003; 2005; 2011b) and the Hyades stars 
(Y{\i}ld{\i}z et al. 2006): {The models for these main-sequence (MS) stars are in good agreement with the seismic and non-seismic 
constraints of these stars.}
2) The second stellar harmonic ($k_2$), required for the apsidal motion analysis of eclipsing binaries, is presented.
3) Fitting formula for effects of rotation on luminosity ($L$), radius ($R$) and $k_2$ for different stellar masses are derived (see below).
4) The grids cover a wide range of stellar mass (0.74 $M_\odot$  to 10.0 $M_\odot$). 
5) The mass step (0.01 $M_\odot$) is so small that properties of a model for a given mass can be found without 
interpolation between models with different masses.
{
6) Updated tables are used for low-temparature opacity (Ferguson et al. 2005).
7) Nuclear reaction rate for $^{14}$N(p,$\gamma$)$^{15}$O is recently updated. We adopt measurement of 
LUNA (Laboratory for Underground Nuclear Astrophysics) 
collaboration for the cross section of this reaction (Bemmerer et al. 2006).
8) Formation of a star depends on how much energy is stored during the contraction process in the pre-MS
 phase.
In this regard, internal structure of zero-age MS (ZAMS) models depends on the details of the pre-MS evolution. 
Therefore, our model computations include pre-MS phase.
9) The {\small ANK\.I} code itself solves Saha equation and computes partition function by using Mihalas et al. (1990) approach
for the surviving probability of atomic/ionic energy levels.
}
The disadvantage of the present grids is that the post-MS phase is not included. The reason for this is that
the {\small ANK\.I} code is not well suited to constructing shell-burning interior models. 
 
Eclipsing binaries with non-circular orbit show apsidal motion. {The period of apsidal motion 
can be found observationally from the O-C analysis of eclipse timings}. Theoretical value of the period is found from the second stellar harmonics,
which are measure of mass distribution in the most outer regions of component stars (see Appendix A).
In the present study, values of $k_2$ are given. $k_2$ is also listed  in Claret (2004).
However, putting aside comparison of the code details, we also give the effect of rotation on $k_2$ as well as luminosity and radius.
This point is important in many fields, such as, for apsidal motion analysis of eclipsing binaries in which rotation influences the structure of their component stars.
We derive fitting formula over certain stellar mass intervals for $k_2$, $L$ and $R$ as a function of rotational parameter $\Lambda$, 
which is basically the ratio of centrifugal acceleration to gravitational acceleration.   
 

The successful application of the grids to the late-type stars and slowly rotating early-type stars may provide 
the mass and age of these types of stars.  
A grid of stellar evolution with a very small mass step is required for such studies. 
Therefore,  we choose the mass step as 0.01 M$_\odot$.
The tables are given for the MS evolution of a given mass and also for isochrones.

 In comparison with non-rotating models with mass pertaining to the early-type stars, the chemically peculiar Am stars, which have a 
very low value of  $v \sin i$ in comparison with their counterparts, are perhaps 
the most suitable ones for the determination of the effectiveness time for diffusion
process. The advantage 
 is that they are slowly rotating A-type stars, in contrast to their spectral type counterparts.
This feature enables the diffusion process to operate. 
The time required for the effectiveness of the 
diffusion process can  be found by using the isochrone fitting method.

Rotation is one of the  essential features of macroscopic objects and, in particular, is very effective on the structure of early-type stars.
Unfortunately, we have very limited information on internal rotation of such stars. While 
the conventional approach 
is to assume solid-body rotation, the alternative is to adopt differential rotation in depth. 
The effect of the former approach on luminosity, radius, and second stellar harmonics as a function of rotational parameter 
is derived by Y{\i}ld{\i}z (2004; see also Stothers 1974).
We also derive similar relations for low-mass stars.
In the case of differential rotation, detailed analysis of 
angular momentum transportation within  stars is required.
This is beyond the scope of the present study.

The remainder of this paper is organized as follows. In Section 2, we present basic properties of the code 
and  initial values required for the model computations.
Section 3 presents tables prepared for the MS evolution of a wide stellar mass interval and for different isochrones.
Section 4 shows some basic results and fitting formula derived from these tables. 
Sections 5 and 6 discuss the comparison of results with
the observations and effect of rotation, respectively. 
In Section 7 we give
concluding remarks.

\section{Basic properties of the code and initial values}
\begin{table}[t]
\caption{Properties of the {\small ANK\.I} code.}
\begin{tabular}{ll}
\hline
Opacity &  OPAL96 \& Ferguson et al. (2005) \\
EOS     &  Saha~ Eq.~ solved \\ 
        &  Coulomb~ interaction~ included  \\
Convection & MLT    \\
Diffusion &  No  \\
Rotation  &  Yes~ \&~ No \\ 
Magnetic Field  & No    \\
\hline
\end{tabular}
\end{table}

{ 
The {\small ANK\.I} code 
used for the present study was first developed in the 1960s by 
D. Eryurt-Ezer and gradually updated by her and her colleagues 
(Ezer \& Cameron 1965; K{\i}z{\i}lo\u{g}lu \& Eryurt-Ezer 1985; Y{\i}ld{\i}z \& K{\i}z{\i}lo\u{g}lu 1997) and more recently by 
Y{\i}ld{\i}z (2000; 2003; 2008).} 
The updated routines 
are for equation of state (EOS), opacity, nuclear reaction rates and chemical advancement due to nuclear 
reactions (see
Table 1 for basic properties of the code). A brief summary is given below.

~~\\
OPACITY - The radiative opacity is derived from
recent OPAL tables (Iglesias \& Rogers 1996; OPAL96 ), implemented by the low temperature tables of {
Ferguson et al. (2005)}. 

~~\\
CHEMICAL COMPOSITION - {X=0.7024 and 
Z=0.0172 values are obtained from calibration of solar models. The present solar surface abundance of heavy elements is reduced to 0.0134 by 
diffusion which is 
in very good agreement with the recent value of 0.0134 found by Asplund et al. (2009). }

~~\\
EOS - 
In the present study, the EOS is obtained by minimization of the free energy
(Mihalas et al. 1990).
Whereas the Saha equation is solved for hydrogen and  helium, ionization degrees of the 
eight most abundant heavy elements (C, N, O, Ne, S, Si, Mg and Fe)
are computed from the expressions given by Gabriel \& Y{\i}ld{\i}z (1995). The basic properties of the routines are described
in detail by Y{\i}ld{\i}z \& K{\i}z{\i}lo\u{g}lu (1997). 

~~\\
NUCLEAR REACTIONS - 
Rate of nuclear reactions are computed from Bemmerer et al. (2006)
and Caughlan \&  Fowler (1988).

~~\\
CONVECTION - Classical mixing-length theory (MLT) of B\"ohm-Vitense (1958) is employed for convection.
While boundaries of convective regions are marked by 
Schwarzschild criteria, overshooting 
is not accounted for. The mixing-length parameter $\alpha =1.98$ is
obtained from by calibrating the solar models.



\begin{table}[t]
\caption{Properties of models.}
\begin{tabular}{lr}
\hline
Initial~ Hydrogen~ abundance~ ($X$) & { 0.7024}  \\
Initial~ heavy~ element~ abundance~ ($Z$) & 0.0172   \\
Mixing-length~ parameter~ ($\alpha$) &  1.98   \\
Minimum~ stellar~ mass  (M$_\odot$) &  0.74  \\
Maximum~ stellar~ mass  (M$_\odot$) &  10.0  \\
Mass~ step  (M$_\odot$) &  0.01  \\
\hline
\end{tabular}
\end{table}

The masses of the models range from 0.74 to 10.0 M$_\odot$. 
The mass step is 0.01 M$_\odot$ through the range. Such a dense grid can be used to estimate 
the mass and age of stars whose spectroscopic and photometric
observations yield data of high quality. Basic properties of models are summarized in Table 2.


\section{Online Tables}
Tables are prepared for grids of stellar evolution, with different masses from ZAMS to terminal-age MS (TAMS), and  
for isochrones using these grids. These tables will appear in the online version of the article. In the following subsections, 
we describe columns of these tables. 
Units of the quantities are in $cgs$, unless specified otherwise.


Although pre-MS phase is included in model computations, to avoid complications, 
only the MS phase from ZAMS to TAMS is presented  in the grids. For isochrones, however, 
pre-MS phase is considered.

ZAMS can be defined as the point at which 
luminosity  or radius  is minimum ($L_{\rm min}$, $R_{\rm min}$). However, for some stellar masses, 
the time difference between the ages from $L_{\rm min}$ and $R_{\rm min}$ is significant.
Alternatively, we determine the ZAMS point for a model 
evolution at which multiplication of luminosity and radius is minimum. 
This is a unique point in most of the evolutionary tracks and therefore very suitable for  automatic
computation
of such large stellar grids.  The TAMS point is adopted as the point at 
which the central hydrogen abundance ($X_{\rm c}$) is 0.0012.  

The following  are given for models with a mass ranging from 0.74 to 10 M$_\odot$. 
The mass step is 
0.01 M$_\odot$. 
\\
1. Column: $\log(t/\rm yr)$   \\
2. Column: $\log(R/{ R_\odot})$   \\
3. Column: $\log(L/{ L_\odot})$   \\
4. Column: $\log(T_{ eff}/\rm K)$   \\
5. Column: $\log(k_{ 2})$   \\
6. Column: $X_{ c}$   \\
7. Column: $(M/{ {\rm M}_\odot})$   \\

Isochrones are obtained for the time interval from $\log (t/\rm yr)=6.90-10.25$,
with the time step $\Delta \log (t/\rm yr)=0.05$.
\\
1. Column: $\log(t/\rm yr)$   \\
2. Column: $\log(R/{ R_\odot})$   \\
3. Column: $\log(L/{ L_\odot})$   \\
4. Column: $\log(T_{ eff}/\rm K)$   \\
5. Column: $M_{V}$   \\
6. Column: $B-V$   \\
7. Column: $U-B$   \\
8. Column: $(M/{ {\rm M}_\odot})$   \\
9. Column: $X_{ c}$   \\

Bolometric correction required for computation of $M_{V}$ is derived from 
Lejeune, Cuisinier, \& Buser (1998). {Colours of the models are also computed by 
interpolation of their tables.}

Sample of online tables are given in Appendix B. 
\section{Results}
   \begin{figure}[htb]
\centerline{\psfig{figure=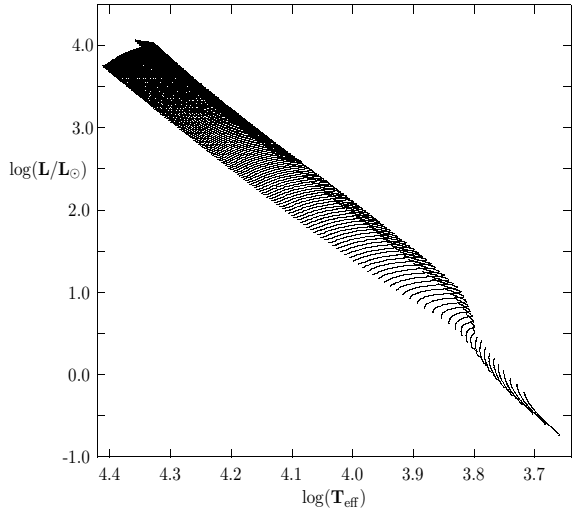,width=180bp,height=210bp}}

\caption{
The HRD for the MS evolutionary tracks of models with solar chemical composition and stellar masses from 0.75 to M=10 M$_\odot$.
}
   \end{figure}

In Fig. 1, the MS evolutionary tracks of models ranging from 0.75 to 10.0 M$_\odot$ with mass step 
0.05 M$_\odot$ are plotted in the Hertzsprung-Russel diagram (HRD). We derive 
and present 
some basic results, which pertain to the ZAMS and TAMS lines and can be useful for astrophysical 
applications. 
Properties of the models with certain masses in ZAMS and TAMS are given in Tables 3 and 4, respectively.

   \begin{figure}[htb]
\centerline{\psfig{figure=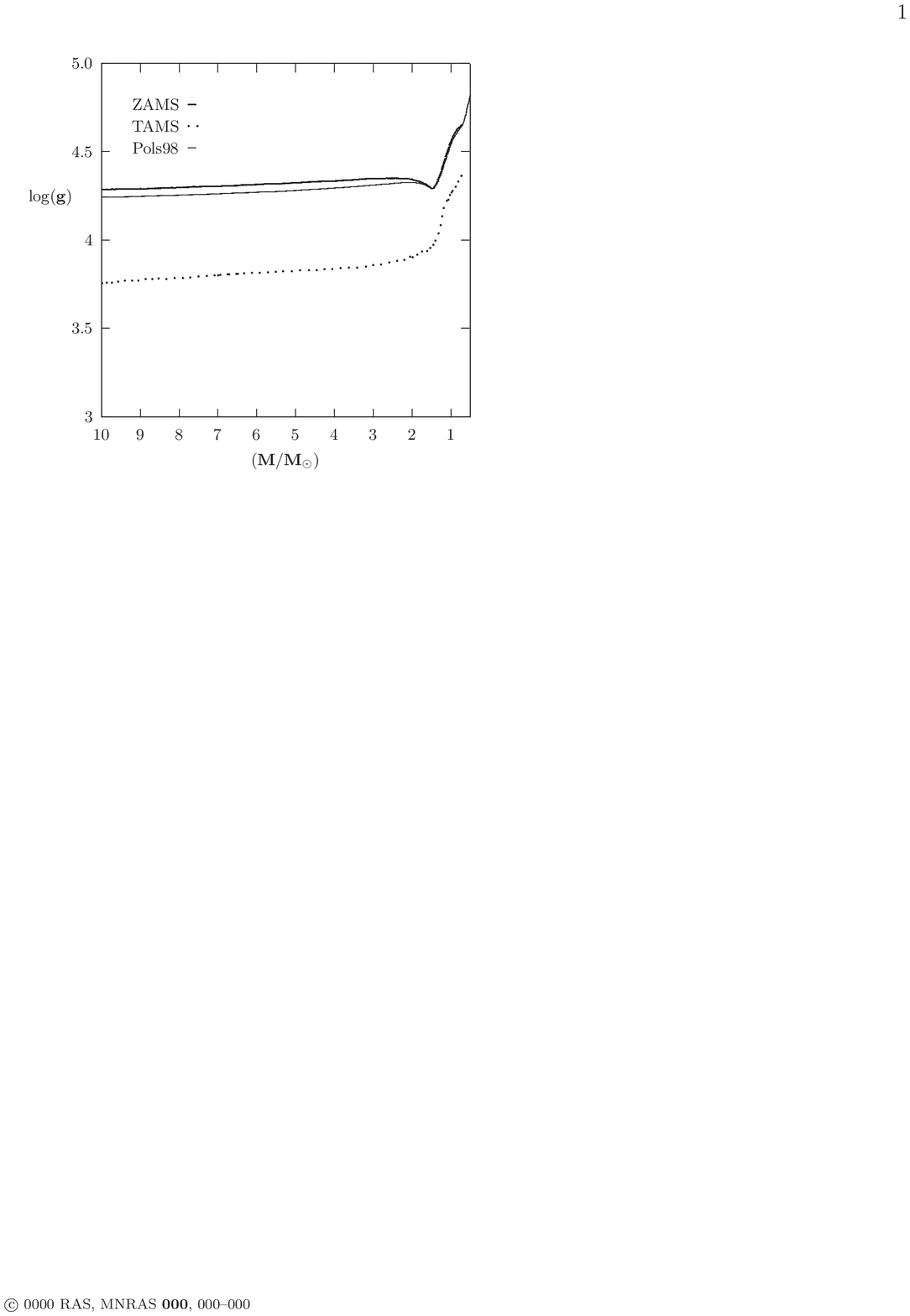,width=250bp,height=225bp }}
\caption{
 $\log(g)$ as a function of stellar mass for the ZAMS (thick solid line) and TAMS 
(dotted line) models.
For comparison, ZAMS values of $\log(g)$ from Pols et al. (1998; thin solid line) are also given.
}
   \end{figure}
   \begin{figure}[htb]
\centerline{\psfig{figure=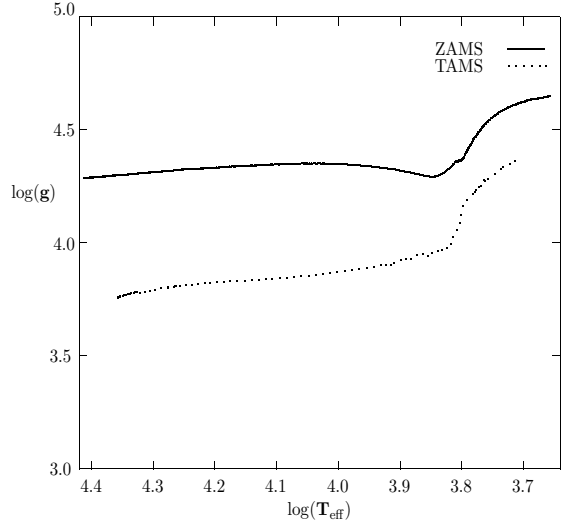,width=180bp,height=210bp}}
\caption{
 $\log(g)$ as a function of effective temperature  for stars at ZAMS and
TAMS.
}
   \end{figure}
\begin{table*}[h]\footnotesize
\caption{
The ZAMS values of some basic parameters of the  models. Mass, luminosity and radius are in solar units.
$\rho_{\rm ph}$ is the density at the surface, while $T_{\rm c}$  and $\rho_{\rm c}$ are the temperature and density at the 
stellar center, in $cgs$. $BC$ represents bolometric correction.
}
$
\renewcommand{\arraystretch}{0.8}
\begin{array}{rrrrrrrrrrrrr}
\hline
           \noalign{\smallskip}
 M & \log R &\log L &\log T_{\rm eff}&\log \rho_{\rm ph}  &\log g &\log k_2 &\log T_c &\log \rho_c &BC~&U-B&B-V&M_{\rm V}\\
\hline
           \noalign{\smallskip}
 0.75  &  -0.167  &  -0.736  &  3.661  &  -6.461  &  4.646  &  -2.440  &  7.048  &  1.889  &  -0.537  &  1.026  &  1.115  &  7.127   \\
 0.80  &  -0.147  &  -0.606  &  3.684  &  -6.539  &  4.634  &  -2.476  &  7.068  &  1.893  &  -0.368  &  0.816  &  0.987  &  6.632    \\
 0.85  &  -0.126  &  -0.485  &  3.704  &  -6.589  &  4.620  &  -2.515  &  7.086  &  1.894  &  -0.273  &  0.587  &  0.883  &  6.235    \\
 0.90  &  -0.104  &  -0.370  &  3.721  &  -6.633  &  4.601  &  -2.558  &  7.102  &  1.894  &  -0.228  &  0.406  &  0.797  &  5.903    \\
 0.95  &  -0.082  &  -0.262  &  3.737  &  -6.672  &  4.579  &  -2.607  &  7.118  &  1.892  &  -0.184  &  0.271  &  0.729  &  5.590    \\
 1.00  &  -0.058  &  -0.160  &  3.751  &  -6.710  &  4.553  &  -2.662  &  7.133  &  1.888  &  -0.148  &  0.156  &  0.669  &  5.299    \\
 1.10  &  -0.006  &  0.028  &  3.772  &  -6.828  &  4.490  &  -2.795  &  7.158  &  1.877  &  -0.110  &  0.042  &  0.588  &  4.790    \\
 1.20  &  0.048  &  0.202  &  3.788  &  -6.962  &  4.420  &  -2.950  &  7.181  &  1.863  &  -0.077  &  -0.001  &  0.532  &  4.323    \\
 1.30  &  0.100  &  0.397  &  3.811  &  -7.171  &  4.352  &  -3.219  &  7.219  &  1.925  &  -0.035  &  -0.037  &  0.456  &  3.792    \\
 1.40  &  0.139  &  0.550  &  3.830  &  -7.347  &  4.307  &  -3.384  &  7.245  &  1.926  &  -0.010  &  -0.034  &  0.400  &  3.386    \\
 1.50  &  0.159  &  0.687  &  3.854  &  -7.559  &  4.295  &  -3.459  &  7.268  &  1.922  &  0.011  &  -0.006  &  0.331  &  3.020    \\
 1.60  &  0.166  &  0.811  &  3.881  &  -7.807  &  4.309  &  -3.453  &  7.287  &  1.911  &  0.019  &  0.060  &  0.261  &  2.704    \\
 1.70  &  0.173  &  0.923  &  3.906  &  -8.037  &  4.322  &  -3.440  &  7.302  &  1.895  &  -0.007  &  0.060  &  0.180  &  2.449    \\
 1.80  &  0.181  &  1.027  &  3.928  &  -8.300  &  4.331  &  -3.428  &  7.315  &  1.877  &  -0.043  &  0.075  &  0.114  &  2.226    \\
 1.90  &  0.189  &  1.124  &  3.948  &  -8.550  &  4.338  &  -3.416  &  7.326  &  1.858  &  -0.093  &  0.058  &  0.066  &  2.034    \\
 2.00  &  0.198  &  1.214  &  3.966  &  -8.767  &  4.343  &  -3.404  &  7.336  &  1.837  &  -0.154  &  0.016  &  0.030  &  1.869    \\
 2.10  &  0.207  &  1.300  &  3.983  &  -8.939  &  4.346  &  -3.393  &  7.344  &  1.816  &  -0.206  &  -0.011  &  0.004  &  1.707    \\
 2.20  &  0.216  &  1.380  &  3.999  &  -9.057  &  4.348  &  -3.382  &  7.352  &  1.794  &  -0.277  &  -0.057  &  -0.018  &  1.576    \\
 2.30  &  0.225  &  1.457  &  4.014  &  -9.129  &  4.350  &  -3.371  &  7.359  &  1.773  &  -0.363  &  -0.115  &  -0.039  &  1.472    \\
 2.40  &  0.234  &  1.530  &  4.027  &  -9.170  &  4.350  &  -3.362  &  7.365  &  1.752  &  -0.447  &  -0.167  &  -0.057  &  1.373    \\
 2.60  &  0.251  &  1.665  &  4.053  &  -9.198  &  4.351  &  -3.343  &  7.376  &  1.712  &  -0.677  &  -0.285  &  -0.093  &  1.263    \\
 2.70  &  0.260  &  1.729  &  4.064  &  -9.197  &  4.350  &  -3.334  &  7.381  &  1.692  &  -0.740  &  -0.321  &  -0.102  &  1.167    \\
 3.00  &  0.284  &  1.906  &  4.096  &  -9.164  &  4.347  &  -3.309  &  7.394  &  1.636  &  -0.780  &  -0.364  &  -0.104  &  0.765    \\
 3.50  &  0.320  &  2.160  &  4.142  &  -9.091  &  4.342  &  -3.272  &  7.413  &  1.553  &  -1.036  &  -0.483  &  -0.130  &  0.385    \\
 4.00  &  0.352  &  2.377  &  4.180  &  -9.033  &  4.336  &  -3.240  &  7.428  &  1.481  &  -1.260  &  -0.568  &  -0.149  &  0.068    \\
 4.50  &  0.380  &  2.564  &  4.213  &  -8.993  &  4.330  &  -3.213  &  7.440  &  1.417  &  -1.454  &  -0.631  &  -0.166  &  -0.206    \\
 5.00  &  0.406  &  2.730  &  4.241  &  -8.965  &  4.325  &  -3.188  &  7.451  &  1.360  &  -1.626  &  -0.682  &  -0.179  &  -0.449    \\
 5.50  &  0.429  &  2.878  &  4.267  &  -8.943  &  4.320  &  -3.167  &  7.461  &  1.309  &  -1.778  &  -0.725  &  -0.190  &  -0.666    \\
 6.00  &  0.451  &  3.011  &  4.289  &  -8.927  &  4.315  &  -3.147  &  7.470  &  1.263  &  -1.911  &  -0.762  &  -0.201  &  -0.865    \\
 6.50  &  0.470  &  3.131  &  4.309  &  -8.913  &  4.310  &  -3.129  &  7.478  &  1.221  &  -2.030  &  -0.795  &  -0.210  &  -1.047    \\
 7.00  &  0.489  &  3.242  &  4.328  &  -8.901  &  4.306  &  -3.113  &  7.485  &  1.183  &  -2.137  &  -0.825  &  -0.218  &  -1.218    \\
 7.50  &  0.506  &  3.344  &  4.345  &  -8.892  &  4.301  &  -3.099  &  7.491  &  1.148  &  -2.234  &  -0.851  &  -0.225  &  -1.376    \\
 8.00  &  0.522  &  3.437  &  4.360  &  -8.886  &  4.298  &  -3.085  &  7.497  &  1.116  &  -2.320  &  -0.875  &  -0.231  &  -1.524    \\
 8.50  &  0.536  &  3.525  &  4.375  &  -8.882  &  4.294  &  -3.073  &  7.503  &  1.086  &  -2.400  &  -0.897  &  -0.237  &  -1.662    \\
 9.00  &  0.550  &  3.606  &  4.388  &  -8.880  &  4.292  &  -3.061  &  7.508  &  1.059  &  -2.474  &  -0.917  &  -0.241  &  -1.791    \\
 9.50  &  0.564  &  3.682  &  4.400  &  -8.880  &  4.288  &  -3.051  &  7.513  &  1.034  &  -2.540  &  -0.935  &  -0.246  &  -1.916    \\
10.00  &  0.576  &  3.754  &  4.412  &  -8.880  &  4.286  &  -3.041  &  7.517  &  1.011  &  -2.604  &  -0.952  &  -0.250  &  -2.030    \\
           \noalign{\smallskip}
            \hline
\end{array}
$

\renewcommand{\arraystretch}{1.0}
\end{table*}
\begin{table*}
\caption{
The TAMS values of some basic parameters of the  models. Mass, luminosity and radius are in solar units.
$\rho_{\rm ph}$ is the density at the surface, while $T_{\rm c}$  and $\rho_{\rm c}$ are the temperature and density at the
stellar center, in $cgs$. $BC$ represents bolometric correction.
}
$
\renewcommand{\arraystretch}{0.8}
\begin{array}{rrrrrrrrrrrrr}
\hline
            \noalign{\smallskip}

 M & \log R &\log L &\log T_{\rm eff}&\log \rho_{\rm ph}  &\log g &\log k_2 &\log T_c &\log \rho_c &BC~&U-B&B-V&M_{\rm V}\\
\hline
 0.75  &  -0.024  &  -0.227  &  3.717  &  -6.756  &  4.361  &  -2.664  &  7.309  &  2.820  &  -0.238  &  0.438  &  0.821  &  5.556   \\
 0.80  &  0.002  &  -0.122  &  3.730  &  -6.793  &  4.336  &  -2.709  &  7.285  &  2.877  &  -0.202  &  0.318  &  0.759  &  5.258    \\
 0.85  &  0.026  &  -0.026  &  3.742  &  -6.827  &  4.315  &  -2.764  &  7.295  &  2.834  &  -0.170  &  0.223  &  0.706  &  4.986    \\
 0.90  &  0.048  &  0.053  &  3.751  &  -6.858  &  4.297  &  -2.812  &  7.282  &  2.810  &  -0.145  &  0.147  &  0.665  &  4.763    \\
 0.95  &  0.069  &  0.131  &  3.760  &  -6.902  &  4.278  &  -2.876  &  7.291  &  2.750  &  -0.126  &  0.089  &  0.628  &  4.548    \\
 1.00  &  0.086  &  0.193  &  3.767  &  -6.940  &  4.265  &  -2.928  &  7.324  &  2.610  &  -0.113  &  0.056  &  0.601  &  4.380    \\
 1.10  &  0.125  &  0.317  &  3.778  &  -7.023  &  4.228  &  -3.036  &  7.331  &  2.468  &  -0.097  &  0.022  &  0.562  &  4.054    \\
 1.20  &  0.170  &  0.472  &  3.795  &  -7.164  &  4.176  &  -3.254  &  7.321  &  2.533  &  -0.061  &  -0.004  &  0.508  &  3.631    \\
 1.30  &  0.238  &  0.655  &  3.807  &  -7.322  &  4.076  &  -3.471  &  7.337  &  2.696  &  -0.038  &  -0.017  &  0.466  &  3.149    \\
 1.40  &  0.290  &  0.808  &  3.818  &  -7.462  &  4.003  &  -3.625  &  7.400  &  2.680  &  -0.018  &  -0.016  &  0.428  &  2.749    \\
 1.50  &  0.323  &  0.931  &  3.833  &  -7.608  &  3.968  &  -3.693  &  7.436  &  2.590  &  0.002  &  -0.004  &  0.383  &  2.421    \\
 1.60  &  0.346  &  1.047  &  3.850  &  -7.777  &  3.950  &  -3.699  &  7.455  &  2.527  &  0.016  &  0.011  &  0.329  &  2.117    \\
 1.70  &  0.363  &  1.151  &  3.868  &  -7.952  &  3.943  &  -3.689  &  7.468  &  2.478  &  0.037  &  0.082  &  0.283  &  1.835    \\
 1.80  &  0.382  &  1.255  &  3.885  &  -8.133  &  3.930  &  -3.683  &  7.477  &  2.444  &  0.029  &  0.108  &  0.226  &  1.584    \\
 1.90  &  0.398  &  1.350  &  3.900  &  -8.318  &  3.920  &  -3.676  &  7.486  &  2.402  &  0.020  &  0.130  &  0.168  &  1.354    \\
 2.00  &  0.416  &  1.446  &  3.915  &  -8.497  &  3.906  &  -3.673  &  7.494  &  2.374  &  -0.004  &  0.129  &  0.112  &  1.138    \\
 2.10  &  0.432  &  1.532  &  3.929  &  -8.688  &  3.897  &  -3.666  &  7.500  &  2.335  &  -0.034  &  0.110  &  0.072  &  0.954    \\
 2.20  &  0.445  &  1.614  &  3.943  &  -8.876  &  3.890  &  -3.658  &  7.507  &  2.306  &  -0.071  &  0.082  &  0.043  &  0.785    \\
 2.30  &  0.456  &  1.690  &  3.956  &  -9.043  &  3.887  &  -3.649  &  7.512  &  2.278  &  -0.115  &  0.043  &  0.015  &  0.639    \\
 2.40  &  0.468  &  1.765  &  3.969  &  -9.186  &  3.883  &  -3.642  &  7.518  &  2.254  &  -0.163  &  0.007  &  -0.009  &  0.499    \\
 2.60  &  0.490  &  1.907  &  3.993  &  -9.372  &  3.873  &  -3.629  &  7.528  &  2.205  &  -0.247  &  -0.056  &  -0.040  &  0.230    \\
 2.70  &  0.501  &  1.973  &  4.005  &  -9.425  &  3.868  &  -3.624  &  7.532  &  2.183  &  -0.311  &  -0.104  &  -0.054  &  0.127    \\
 3.00  &  0.528  &  2.158  &  4.037  &  -9.483  &  3.858  &  -3.606  &  7.545  &  2.123  &  -0.505  &  -0.233  &  -0.088  &  -0.140    \\
 3.50  &  0.568  &  2.422  &  4.083  &  -9.458  &  3.845  &  -3.579  &  7.563  &  2.031  &  -0.694  &  -0.354  &  -0.109  &  -0.611    \\
 4.00  &  0.601  &  2.648  &  4.123  &  -9.400  &  3.838  &  -3.557  &  7.579  &  1.959  &  -0.928  &  -0.467  &  -0.132  &  -0.942    \\
 4.50  &  0.630  &  2.841  &  4.157  &  -9.351  &  3.831  &  -3.536  &  7.592  &  1.883  &  -1.122  &  -0.544  &  -0.148  &  -1.231    \\
 5.00  &  0.655  &  3.012  &  4.187  &  -9.316  &  3.826  &  -3.518  &  7.605  &  1.824  &  -1.299  &  -0.605  &  -0.163  &  -1.481    \\
 5.50  &  0.679  &  3.165  &  4.213  &  -9.292  &  3.820  &  -3.504  &  7.617  &  1.773  &  -1.455  &  -0.653  &  -0.175  &  -1.707    \\
 6.00  &  0.701  &  3.302  &  4.237  &  -9.276  &  3.814  &  -3.492  &  7.627  &  1.729  &  -1.593  &  -0.695  &  -0.186  &  -1.911    \\
 6.51  &  0.721  &  3.428  &  4.258  &  -9.263  &  3.809  &  -3.481  &  7.637  &  1.683  &  -1.719  &  -0.733  &  -0.195  &  -2.101    \\
 6.95  &  0.737  &  3.529  &  4.275  &  -9.255  &  3.805  &  -3.475  &  7.646  &  1.659  &  -1.819  &  -0.761  &  -0.203  &  -2.255    \\
 7.50  &  0.759  &  3.645  &  4.293  &  -9.251  &  3.795  &  -3.470  &  7.654  &  1.612  &  -1.922  &  -0.792  &  -0.211  &  -2.440    \\
 8.00  &  0.777  &  3.742  &  4.309  &  -9.249  &  3.787  &  -3.466  &  7.662  &  1.582  &  -2.009  &  -0.818  &  -0.217  &  -2.595    \\
 8.50  &  0.794  &  3.832  &  4.323  &  -9.249  &  3.779  &  -3.464  &  7.668  &  1.548  &  -2.086  &  -0.841  &  -0.223  &  -2.743    \\
 9.00  &  0.809  &  3.915  &  4.336  &  -9.246  &  3.775  &  -3.461  &  7.677  &  1.528  &  -2.162  &  -0.863  &  -0.228  &  -2.876    \\
 9.50  &  0.825  &  3.996  &  4.348  &  -9.251  &  3.764  &  -3.463  &  7.684  &  1.506  &  -2.226  &  -0.883  &  -0.232  &  -3.014    \\
10.00  &  0.841  &  4.068  &  4.358  &  -9.257  &  3.756  &  -3.462  &  7.688  &  1.478  &  -2.282  &  -0.900  &  -0.235  &  -3.140    \\

            \noalign{\smallskip}
            \hline
\end{array}
\renewcommand{\arraystretch}{0.8}
$
\end{table*}

ZAMS age as a function of stellar mass is found as 
\begin{equation}
t_{\rm ZAMS}=\frac{8.05 \times 10^7}{(M/{\rm M}_\odot)^{2.22}}\rm yr.
\end{equation}
{The maximum difference between age from equation (1) and model age is about 20 per cent.}
Very precise TAMS age as a function of stellar mass for the same mass interval is derived as
\begin{equation}
t_{\rm TAMS}=\frac{10^{10}}{(M/{\rm M}_\odot)^{4.05}}(5.60\times 10^{-3}(\frac{M}{{\rm M}_\odot}+3.993)^{3.16}+0.042)~\rm yr. 
\end{equation}
{The accuracy of equation (2) is very high for the models with $M>2$ M$_{\sun}$. The maximum difference between 
its prediction and model age is 5 per cent, for this mass range. It is about 15 per cent for the range $M<2$ M$_{\sun}$.
}

%

In Fig. 2,  $\log(g)$ of stars at ZAMS (thick solid line) and TAMS (dotted line) is 
 plotted with respect to 
stellar mass. For comparison, $\log(g)$ of ZAMS models (thin solid line) of Pols et al. (1998; Pols98) is also plotted. 
The dependence of $\log(g)$ on stellar mass is very different for models with $M>$ 1.5 M$_\odot$ and $M<$ 1.5 M$_\odot$. 
For models of the early-type stars ($M>$ 1.5 M$_\odot$), both ZAMS and TAMS values of $\log(g)$ are nearly constant, and are about 
4.3 and 3.8, respectively, whereas
$\log(g)$ of late-type stars ($M<$ 1.5 M$_\odot$) is 
much more sensitive function of stellar mass 
than that of the early-type stars. As stellar mass reduces, $\log(g)$ increases 
and reaches values 4.7 for ZAMS and 
4.4 for TAMS of 0.74 M$_\odot$ model. 
The difference between $\log(g)$ values of ZAMS and TAMS is 0.5 for the early-type stars 
and 0.3 for the late-type stars.
In Fig. 3, $\log(g)$ 
is also plotted with respect to $T_{\rm eff}$. On this plot, the characteristics of $\log(g)$ changes at about $T_{\rm eff}$=6750 K.


\subsection{Results on early-type stars ($M>$ 1.5 M$_\odot$) }
The central regions of these stars have relatively high temperatures.
For these stars, the nuclear reactions proceed via the CNO cycle, which is 
much more productive than the proton-proton chain. 
Therefore, one should expect different MS lifetime values for early- and late-type stars. For the former,
we derive 
\begin{equation}
t_{\rm TAMS}=\frac{4.50 \times 10^9}{(M/{\rm M}_\odot)^{2.40}} \rm yr.
\end{equation}
This expression (see also equation 7) is simpler but less accurate than equation (2).

In Fig. 1, luminosity is minimum at ZAMS and gradually increases during MS evolution for all the models.
$L_{\rm TAMS}$ = 2 $L_{\rm ZAMS}$ is a very good approximation for the full mass range.
The mass-luminosity relation for TAMS of 
early-type stars is as follows 
\begin{equation}
\frac{L_{\rm TAMS}}{L_{\rm \odot}}={2.22}{(M/{\rm M}_\odot)^{3.77}}.
\end{equation}
{The maximum difference between
prediction of equation (4) and model luminosity is about 10 per cent, for this mass range. It is about 15 per cent 
for the range $M<2$ M$_{\sun}$ (see the text below Eq. (7) in Section 4.2).
}

We also derive a mass-effective temperature relation for TAMS and ZAMS:
\begin{equation}
\frac{T_{\rm eff,TAMS}}{T_{\rm eff\odot}}=-0.1237\left(\frac{M}{{\rm M}_\odot}+0.4831\right)^{1.5}+0.7795\frac{M}{{\rm M}_\odot}
+0.3496,
\end{equation}
\begin{equation}
\frac{T_{\rm eff,ZAMS}}{T_{\rm eff\odot}}= 2.3751\left(\frac{M}{{\rm M}_\odot}-0.12\right)^{0.4}-1.4597,
\end{equation}
{where effective temperature of the Sun ($T_{\rm eff\odot}$) is taken as 5777 K.
Equations (5) and (6) are useful for analysis of observed data. 
 Uncertainties in equations (5) and (6) 
are 40 and 50 K, respectively.
}



\subsection{Results on late-type stars ($M<$ 1.5 M$_\odot$)}
We derive expression for the TAMS ages of late-type stars as  
\begin{equation}
t_{\rm TAMS}=\frac{9.65 \times 10^9}{(M/{\rm M}_\odot)^{3.94}} \rm yr.
\end{equation}
{
For TAMS luminosity of the late-type stars, we  find 
}
\begin{equation}
\frac{L_{\rm TAMS}}{L_{\rm \odot}}={1.64}{(M/{\rm M}_\odot)^{3.81}}.
\end{equation}
{In comparison with model luminosities, equation (8) is uncertain about 15 per cent.
}

{
The fitting formula for ZAMS and TAMS $T_{\rm eff}$ of late-type stars as a function of stellar mass is found as  }
\begin{equation}
\frac{T_{\rm eff,ZAMS}}{T_{\rm eff\odot}}=0.333\left(\frac{M}{{\rm M}_\odot}-0.7041\right)^{0.4}+0.2817\frac{M}{{\rm M}_\odot} +0.4835,
\end{equation}
\begin{equation}
\frac{T_{\rm eff,TAMS}}{T_{\rm eff\odot}}=-0.4689\left(\frac{M}{{\rm M}_\odot}+2.2635\right)^{1.5}+1.6442\frac{M}{{\rm M}_\odot}
+2.1294.
\end{equation}
For early-type stars and the hot side of late-type stars in HRD, $T_{\rm eff,ZAMS}$ is higher than $T_{\rm eff,TAMS}$. However, for models with 
$M<1.29$ M$_\odot$,
$T_{\rm eff,TAMS}>T_{\rm eff,ZAMS}$.
{
Uncertainties in equations (9) and (10) are  
60 and 40 K, respectively.
}

Some of the fitting formula derived in this section are plotted in Figs. C.1-3 of Appendix C.

\subsection{Depth of convective zones in late-type stars}
   \begin{figure}[htb]
\centerline{\psfig{figure=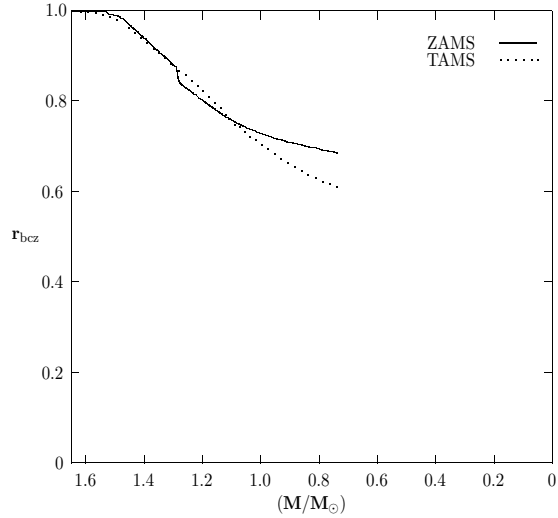,width=180bp,height=210bp}}
\caption{
 The base radius of convective zones in late-type stars as a function of stellar mass.
}
   \end{figure}


In the outer regions of late-type stars, opacity is so high that the radiative temperature gradient is greater than 
the adiabatic gradient.  
It is usually reasoned that the stars with mass less than 0.5 M$_\odot$ have such an opaque interior that they 
are completely convective
(see, for example, Mullan \& MacDonald 2001; Browning 2008; Morin et al. 2008). 
This reasoning is very interesting in two respects: 1) Could convection mix the rare envelope and dense core in these stars?
2) If yes, do these stars ignite all their hydrogen as nuclear fuel?

In late-type star models, density is so high that the  
non-ideal effects should be taken into account. In these models, the assumption of ideal gas pressure (including degeneracy) 
may not be justified. Hence, Coulomb interaction should be considered.
For low-mass stars, in some regions, coulomb energy becomes comparable to the thermal kinetic energy.
In such a case, expression for coulomb energy such as given in Landau and Lifshitz (1969) is no longer valid. 


The convective zone deepens as stellar mass decreases.
This phenomenon is depicted in Fig. 4 which displays the base radius of the convective zone, in the unit of total model radius 
($r_{\rm bcz}=R_{\rm bcz}/R_\star$) as a function of stellar mass.
The dotted line is 
for TAMS and solid line is for ZAMS.
{ The fitting curve for $r_{\rm bcz}$ of TAMS
is given as $0.20 (M/M_\odot)^{2.4}+0.51$. 
}



\subsection{Ionization degree of the most abundant chemical elements at stellar surfaces}
   \begin{figure}[htb]
\centerline{\psfig{figure=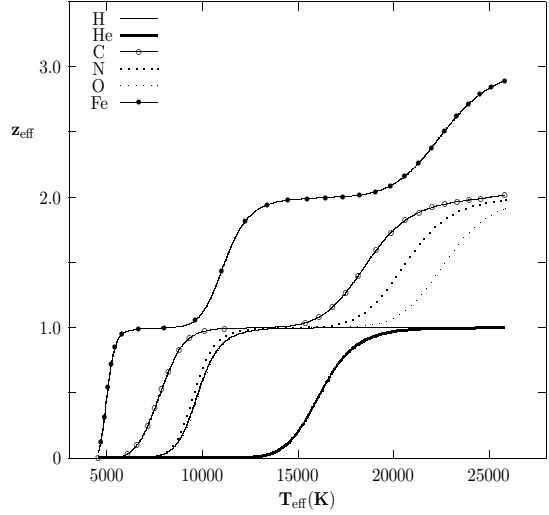,width=180bp,height=210bp}}
\caption{
 Ionization degree of the most abundant chemical elements at the surface of stellar models (ZAMS) as a function of effective temperature.
The thin and thick solid lines are for H and He, respectively. The thin and thick dotted lines are for O and N, respectively.
The open circle represents C, while the filled circles are for Fe.
}
   \end{figure}

The spectral class of a star is determined from its spectral lines.
Absorption lines observed in spectra of stars are mainly due to the transition of atomic and ionic electrons from
one bounded quantum state to another. These lines are extremely important to decipher the properties of stars.
The observability of a line pertaining to a certain type of ion primarily depends on how abundant this ion in the observed medium is.
As a matter of fact, excitation paves the way for ionization.
In Fig. 5, the effective charges (mean ionization degree, $z_{\rm eff}$) of the most abundant elements with respect to effective temperature are plotted.
The thin solid line represents hydrogen. Ionization of hydrogen starts at about $T_{\rm eff}$=7500 K and ends at about $T_{\rm eff}$=12500 K. The curves of H, N (thick dotted line) 
and O (thin dotted line) are very close to each other for the stars 
having $T_{\rm eff}$ less than 15000 K, because first ionization potentials of these elements are nearly the same. For stars with $T_{\rm eff}$ around 25000 K, C ( open circle), N and
O have lost two electrons. However, He (thick solid line) is singly ionized at such a high effective temperature. 
The most ionized element among those plotted in Fig. 5 is Fe ( filled circle). 
{
Single ionization of Fe starts at $T_{\rm eff}$ slightly less than 5000 K 
and ends at $T_{\rm eff}$ slightly higher than 5000 K. Thus, absorption lines of neutral (Fe {\small I}) and first ionized (Fe {\small II}) iron appear about 5000 K. 
Transition from Fe {\small II} to Fe {\small III} occurs at about 10000 K. 
Fe {\small IV} starts to appear at about 20000 K. 
For $T_{\rm eff}$ about 25000 K, Fe is three times ionized.
In stars with  $T_{\rm eff}$ higher than 25000 K, no line of 
Fe {\small IV} (Fe$^{+++}$) is observable.    
}

\section{Comparison of results with observational constraints and other studies}
\subsection{Comparison of solar models }

   \begin{figure}[htb]
\centerline{\psfig{figure=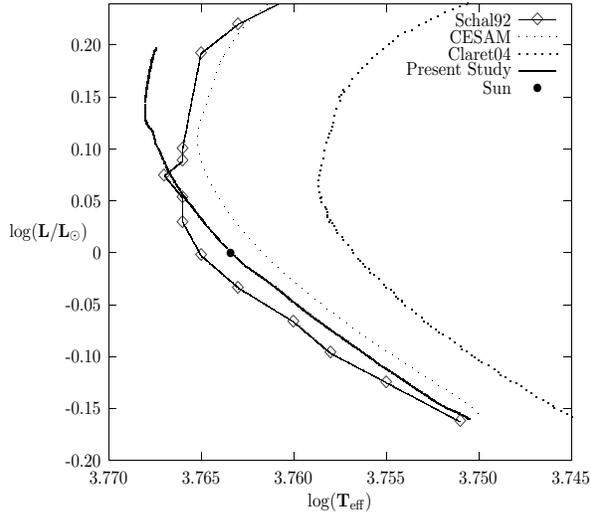,width=180bp,height=210bp}}
\caption{
 Comparison of 1 M$_\odot$ models in HRD.
}
   \end{figure}
The Sun is the nearest star to us and is always the first object to model for a stellar evolution code. 
The precise seismic and non-seismic constraints to the solar interior are very important for
our understanding of stellar structure and evolution. 
In most cases, solar values for chemical composition and convective parameter 
are used for stars if there is no constraint for them.
 
In Fig. 6, the results from the model of 1 M$_\odot$ are plotted in the HRD. For comparison, the results from other studies
and the Sun are also shown. The best agreement is achieved by the present study, despite the 
fact that the diffusion process is not included.
{The differences are small, but it seems that Schaller et al. (1992; Schal92) and CESAM's 
(Code d'Evolution Stellaire Adaptatif et Modulaire; Lebreton \& Michel 2008) 
codes are not well calibrated on the solar parameters. 
The evolutionary track of 
Claret (2004; Claret04), on the other hand, substantially deviates from the data for the Sun. 
He finds X=0.684 and Y=0.296 from the calibration of solar model but the tracks are tabulated for X=0.7, Y=0.28.
The disagreement is due to the fact that the chemical composition of Claret's grids is not the 
same as the solar composition.
}
\subsection{Comparison for Hyades cluster }
   \begin{figure}[htb]
\centerline{\psfig{figure=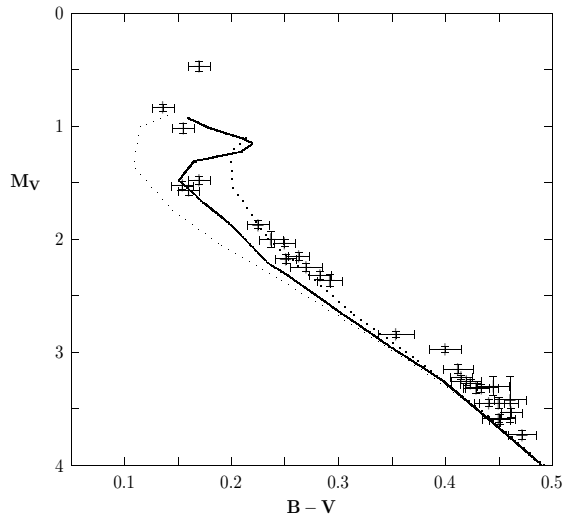,width=180bp,height=210bp}}
\caption{
Colour-magnitude diagram for the Hyades stars. The lines show the isochrones.
The solid line is for the age $\log(t/{\rm yr})=8.90$. The thin and thick dotted lines are for,
 the age $\log(t/{\rm yr})=8.85$ and $\log(t/{\rm yr})=8.95$, respectively.
}
   \end{figure}

Star clusters are the test objects of astrophysics in many respects and 
 are the main laboratory for stellar structure and evolution 
in the early phase of stellar astrophysics. 
The advantage of studying clusters is that their members
are assumed to be formed from the same material and at the same time. 
Hyades is the nearest cluster to us and therefore 
its members are among the most precise distances and hence the absolute magnitudes.

{In Fig. 7}, the Hyades stars with very precise observational data (de Bruijne et al. 2001) are plotted 
in the colour-magnitude diagram (CMD). Also shown are three isochrones. 
The solid line is for $\log(t/{\rm yr})= 8.90$.
The thin and thick dotted lines, however, are for $\log(t/{\rm yr})= 8.85$ and $\log(t/{\rm yr})= 8.95$, respectively.
{
For $B-V>0.3$, the three isochrones are almost equivalent.
They  draw border of the hotter side of MS. 
This is reasonable because the isochrones 
are produced from non-rotating models and rotation moves position of the data toward the red side of the 
 CMD. Binarity also causes upward and rightward shifts in CMD . Therefore,
for agreement between isochrones and the observed data, the isochrone line must be in the blue side of the data.
The isochrone for $\log(t/{\rm yr})= 8.85$ is not in agreement with the brightest stars.
A similar situation is also valid for $\log(t/{\rm yr})= 8.95$.  
The best fitting isochrone is for $\log(t/{\rm yr})= 8.90$ ($t=794$ Myr).
}
This age is  in agreement with 720 Myr found by Y{\i}ld{\i}z et al. (2006) from the 
binaries of Hyades and 
the value ($\log(t/{\rm yr})= 8.896$) given by the WEBDA database (Mermilliod 1995; www.univie.ac.at/webda).

\subsection{Comparison for Am binaries}
   \begin{figure}[htb]
\centerline{\psfig{figure=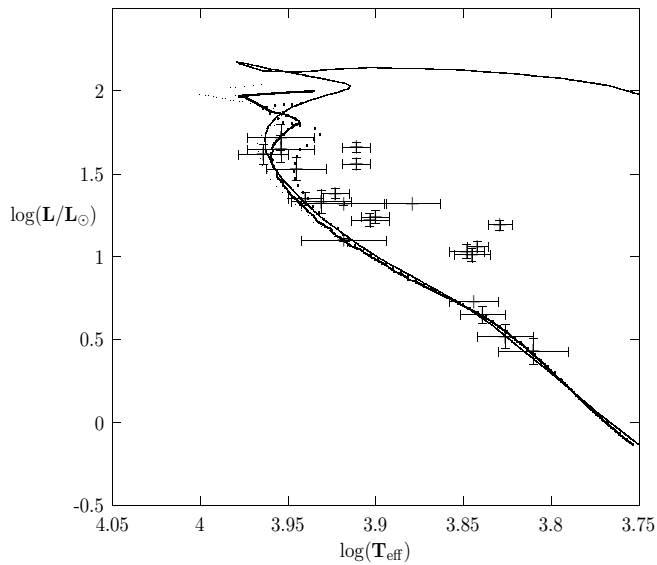,width=200bp,height=231bp}}
\caption{
The theoretical HRD is plotted for the Am stars in double-lined eclipsing binaries.
Also shown the isochrones at ages $t=355$ (thin dotted line), $447$ (thick solid line) and $562$ (thick dotted line) Myr. The thin solid line is for the isochrone at age 450 Myr given by Salasnich et al. (2000).
}
   \end{figure}
Diffusion is one of the microscopic processes operating inside the stars which in turn 
changes 
their observable quantities. For the sun for example, the sound speed difference between the Sun and 
the solar models is reduced if diffusion of  helium and heavy elements is included in the model computations.
Some early-type stars have so high abundance of certain elements that only the diffusion process
can fulfill. The diffusion process in early-type stars is a slow process but fast enough to 
change spectral properties of stars, provided that
rotational velocity is slow. The non-magnetic chemically peculiar Am stars are slow rotators and therefore
very appropriate early-type stars for confrontation with non-rotating stellar models.   
 
{In Fig. 8}, the theoretical HRD is plotted for  the Am stars in
double-lined eclipsing binaries (Andersen 1991). { Also shown are the isochrones at ages 355 (thin dotted line), 447 
(thick solid line), and 562 Myr (thick dotted line).
The isochrone of 447 Myr is in very good agreement with the observational properties of  the Am stars.
For comparison, the isochrone given by  Salasnich et al. (2000) for the same age is also plotted in Fig. 8 
(thin solid line). Two isochrones with the same age is in agreement.
This age is the time required for the microscopic diffusion process to be effective.  
That is to say, during this time interval, metals such as Zn and Sr are gradually levitated. As a result of the levitation, 
after about 450$\pm$100 Myr, photosphere of A-type stars with low rotational velocity abounds in such elements.
\subsection{Comparison of apsidal advance }
   \begin{figure}[htb]
\centerline{\psfig{figure=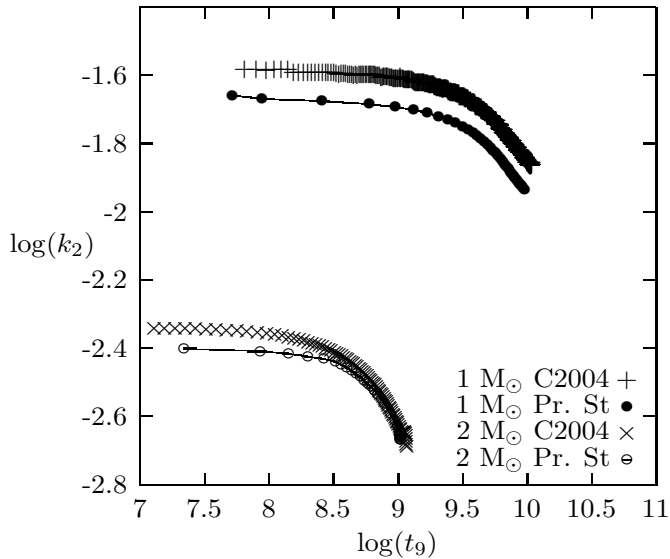,width=250bp,height=225bp}}
\caption{
$k_2$ of 1 M$_\odot$ (filled circle) and 2 M$_\odot$ (open circle) models are plotted with respect to age.
For comparison, $k_2$ given by Claret (2004; C2004) is also plotted.
}
   \end{figure}
   \begin{figure}[htb]
\centerline{\psfig{figure=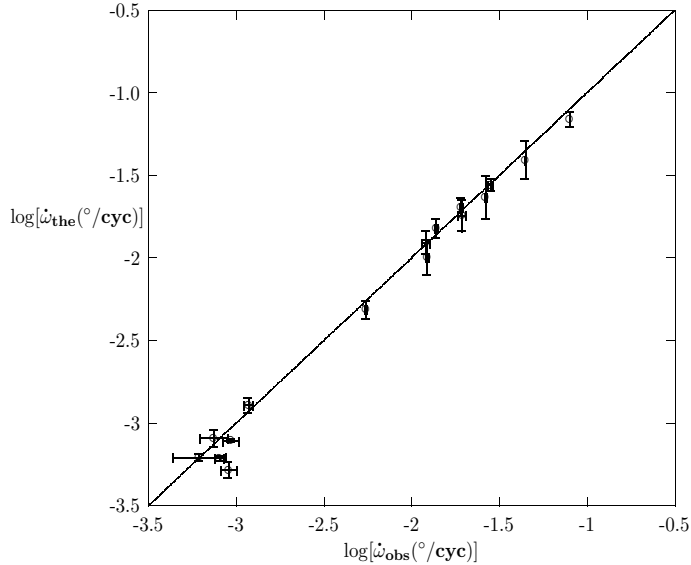,width=200bp,height=230bp}}
\caption{
 Comparison of theoretical and observational apsidal advance rates.
}
   \end{figure}
The existence of many indicators resulting from physical processes occurring inside the stars leads 
us to discover the internal structure and evolution of stars. Three of these indicators are classical 
diagnostics of stellar structure and permit us to see inside the stars. These are the 
detection of 
neutrinos yielded via nuclear reactions in the central regions, frequencies of asteroseismic oscillations trapped in the cavities inside the stars and 
apsidal motion. The last of these is observed in the eccentric eclipsing binaries.
The observed apsidal motion rate is computed using the timing of the changing position of the eclipses. 
The theoretical rate, however, can be expressed in terms of the second harmonic ($k_2$) of the component 
stars. For some binary systems, general relativistic effect must be taken into account. $k_2$ is a measure 
of mass distribution in the most outer regions of component stars.

In literature, there are many studies of many investigators on apsidal motion of eclipsing binaries 
(see Zasche 2012; Wolf et al. 2010; Claret and Gim\'{e}nez 2010; Bulut 2009; Bak{\i}\c{s} et al. 2008;
Khaliullin and Khaliullina 2007; Wolf et al. 2006).
Some well known binaries have components which are so close, but not contacting, that the time coverage of their eclipse 
data is comparable to their apsidal motion period. Binaries with precise apsidal motion period 
are the most suitable ones for apsidal motion analysis. 

{
$k_2$ is plotted with respect to $t_9=t/10^9$ yr in Fig. 9 for the models with 1 M$_\odot$ and 2 M$_\odot$.
As a star evolves in MS, its outer regions expand while the central regions contract. Therefore, $k_2$, a measure of mass 
distribution in stellar envelope (see equation A1), decreases with time during the MS phase.
For comparison, $k_2$ given by Claret (2004; C2004) is also shown in Fig. 9. There is a systematic 
difference between $k_2$ of models with 1 M$_\odot$. It is about $\Delta \log k_2\approx 0.08$.
The initial values of $X$, $Z$ and $\alpha$ in our and Claret's models are quite different. 
The difference $\Delta \log k_2\approx 0.08$ may arise from usage of different initial values.
For the 2 M$_\odot$ models, there is an agreement as the models evolve toward the TAMS.
}

The apsidal motion of the eclipsing binaries with well known accurate dimensions have 
been recently studied by 
Claret and Gim\'{e}nez (2010). In order to compare observational and  theoretical apsidal advances
for well known binaries given in Claret and Gim\'{e}nez (2010) and Claret and Williems (2002),
we compute the apsidal advance rate ($\dot{\omega}_{the}$) from $k_2$ of the present stellar grids
(see Appendix A). 

{We first compute the age of a binary system using the mass and radius of its primary component.
We find the time ($t$) at which the model radius is equal to the observed radius.
Theoretical apsidal advance rate $\dot{\omega}_{the}$ is computed from $k_2$ of component stars at $t$.}
{In Fig. 10}, $\dot{\omega}_{the}$ is plotted with respect to $\dot{\omega}_{obs}$. 
Method for computation of uncertainty in $\dot{\omega}_{the}$ is given in Appendix A.
In such a logarithmic graph,
the theoretical and observational advance rates are in very good agreement. However, for some binaries 
with short period apsidal motion the uncertainties are significantly less than the difference between
$\dot{\omega}_{the}$ and $\dot{\omega}_{obs}$. Therefore, such binaries need further detailed analysis. 
As an example of apsidal motion inference, internal rotation of components of PV Cas has been studied by the 
author of the present paper (Y{\i}ld{\i}z 2005).  

}
\section{Effect of Rotation}
\begin{table*}
\small
\caption{
The effect of rotation for a given surface value of rotational parameter depends on the stellar mass.
The parameters for the effect of rotation on luminosity, radius, and the second stellar harmonic are listed}
\begin{tabular}{llllccllllllll}
 \hline\noalign{\smallskip}
$M_{\rm }$/M$_\odot$& $c_L$   & $c_R$   & $c_{k2}$ \\
 \hline\noalign{\smallskip}
1.0   &   0.760 & 0.225& -0.097   \\  
1.2   &   0.484 & 0.250& -0.058    \\  
1.4   &   0.221 & 0.357& -0.342    \\  
2.0   &   0.184 & 0.507& -0.767   \\  %
2.4   &   0.215 & 0.465& -0.752  \\  %
2.8   &   0.250 & 0.450 & -0.700   \\  
 \hline\noalign{\smallskip}
\end{tabular}
\end{table*}

Apart from the chemically peculiar (Ap Bp and Am) stars, the early-type stars are rapid rotators reaching $v \sin(i)=300$ km s$^{-1}$.
Therefore, the effect of rotation must be included for the early-type stars at least. The rotational velocity directly derived from spectra of stars is the 
velocity of their photosphere. However, for an exhaustive rotating model,
a complete knowledge of internal rotation is required. 
{Recently, diferentially rotating models are constructed by Ekstr\"om et al. (2012). They assume 
solid-body rotation  at ZAMS and later they allow differential rotation. 
}
In the  previous literatures, a solid-body rotation is widely assumed for the representation of internal rotation. However, this assumption does not hold for some early-type stars
(Y{\i}ld{\i}z 2003; 2005). 

In the simple case of solid-body rotation, the stellar parameters could be 
derived in terms of rotational parameter 
$\Lambda_r$, which is defined as 
\begin{equation}
\Lambda_r=\frac{2\Omega^2 r^3 }{3GM(r)},
\end{equation}
where $M(r)$, $G$ and $\Omega$ are the mass inside the sphere with radius $r$,
the universal gravitational constant and rotational angular velocity at $r$, respectively.
{$\Lambda_r$ is very small near the central regions and is maximum in the most outer regions.
Rotation influences the hydrostatic structure of stars in an amount depending upon value of $\Lambda_r$.
In the case of solid-body rotation, the most influnced part is the most outer regions and therefore 
the largest difference between rotating and non-rotating models occurs in their radii. 
The more rapid rotation is, the higher the radius is. 
For luminosity, however, the situation is different. Rotation causes formation of nuclear core cooler than
that of the non-rotating counterpart of a model. Therefore, the higher the rotation rate is, 
the lower the luminosity is.

Luminosity of 
a rotating model can be expressed in terms of
the luminosity of its non-rotating counterpart ($L_{\rm o}$) and the value of the 
rotational parameter ($\Lambda_{\rm s}$) at the surface (Y{\i}ld{\i}z 2005):
} 
\begin{equation}
L=\frac{L_{\rm o}}{(1+\Lambda_{\rm s})^{0.25}}.
\end{equation}
The effect of solid-body rotation on the radius, on the other hand, can be formulated from the model properties as
\begin{equation}
R={R_{\rm o}}{(1+\Lambda_{\rm s})^{0.45}}.
\end{equation}
Rotation also influences stellar harmonic $k_{\rm 2}$. In terms of $\Lambda_{\rm s}$, the change can be written as
\begin{equation}
\Delta \log k_{\rm 2}=-0.7 \Lambda_{\rm s},
\end{equation}
{where $ \Delta \log k_{2}$ is the logarithmic difference between $k_{\rm 2}$ of rotating and non-rotating models,
$\Delta \log k_{2}=\log k_{2}(\Omega)-\log k_{2}(\Omega=0)$.
}

The effect of rotation given in equations (12)-(14) are not valid for all the mass range. 
Therefore, we have constructed rotating models for the masses 1.0, 1.2, 1.4, 2.0, and 2.4 M$_\odot$.
From these rotating models, we derive equations similar to equations (12)-(14) with $c_L$, $c_R$, and $c_{k2}$
defined as 
\begin{equation}
L_{\rm rot}=\frac{L_{\rm o}}{(1+\Lambda_{\rm s})^{c_L}},
\end{equation}
\begin{equation}
R_{\rm rot}={R_{\rm o}}{(1+\Lambda_{\rm s})^{c_R}},
\end{equation}
\begin{equation}
\Delta \log k_{\rm 2}=-c_{k2} \Lambda_{\rm s}.
\end{equation}
The coefficients $c_L$, $c_R$, and $c_{k2}$ are listed  in Table 5. 

Rotating models constructed by using the ANK\.I code are compared with that of other studies in the literatures.
If we compare luminosities, our one-dimensional models (see Fig. 4 in Y\i ld\i z 2004) are in very good agreement with 2-dimensional models obtained by Roxburgh (2004). The radii obtained by Roxburgh are also in good agreement with the 
radii we obtained, provided that we take mean radius of a 2-dimensional model as the geometrical mean of
equatorial and polar radii. Our results concerning $k_2$ of rotating models are  in good agreement with results of Stothers (1974) given for early-type stars.

\section{Conclusions}
Starting from threshold of stability point at which gravitational and internal 
energies are nearly the same, we construct a series of evolutionary models for the mass interval
of 0.74-10.0 M$_\odot$, with a mass step of 0.01 M$_\odot$. 
The results are presented as grids of stellar evolution and isochrones. We derive some  basic expressions for ages, luminosities, and 
effective temperatures for ZAMS and TAMS which may be useful for astrophysical applications.
We also obtain some expressions for certain stellar masses about how rotation affects the fundamental properties of 
MS stars.

We also discuss how deep convective envelope of the coolest stars may be. It seems that there are no full 
convective stars and the maximum size of the convective envelope is about half the star radius. 

We compare model results with the observational results of the Sun, the Hyades cluster, 
the chemically peculiar Am stars and the eclipsing binaries with apsidal motion, and confirm a good agreement
between the results.


\begin{acknowledgements}
Cenk Kayhan and \.Ilknur Gezer are  acknowledged for checking the language of the manuscript.
This work is supported by the Scientific and
Technological Research Council of Turkey (T\"UB\.ITAK 112T989).
\end{acknowledgements}

\newpage
\onecolumn
\appendix
\section[t]{Apsidal Advance}
\subsection[t]{Apsidal Advance in terms of stellar harmonics}

{
Although the apsidal motion is active in all binary systems, we can measure it only 
in the eclipsing binary (or triple) systems with elliptic orbit. Due to tidal interaction
and rotational flattening, the stars are not perfectly spherical. In this case, the second stellar
harmonic of component $i$ (1 for primary and 2 for secondary component) as a measure of 
mass distribution throughout the star (Martynov 1973) is given as
\begin{equation}
k_{2i}=\frac{16\pi }{5} \int_{0}^{R_i} \frac{\rho_i(r)}{M_i(r)}
\left( \frac{r}{R_i}\right) ^{5} r^2 dr
\end{equation}
where $r$ is the radius of the sphere having a mass of $M_i(r)$, $\rho_i(r)$ is the mass 
density at radius $r$, and, $R_i$ and $M_i=M_i(R_i)$ are 
the radius and the total mass of the component $i$, respectively. Since $(r/R_i)^{5}$ is 
negligible in the central region, $k_{2i}$ depends mostly on the mass distribution in outer 
regions. The apsidal motion is determined by $k_{2i}$.
Angular velocity of apsidal advance in the direction of
orbital motion  is given by Kopal (1978) as
\begin{eqnarray}
\dot{\omega}_{cl}&=&k_{21} b_1^5 \left[15 f_2(e)\frac{M_2}{M_1}+\left(\frac{\omega_{r,1}}
                    {\omega_{k}}\right)^2 \left(1+\frac{M_2}{M_1}\right)g(e)\right]  \\ 
       &+&k_{22} b_2^5 \left[15 f_2(e)\frac{M_1}{M_2}+\left(\frac{\omega_{r,2}} {\omega_{k}} \right)^2 \left(1+\frac{M_1}{M_2}\right)g(e)\right]\nonumber\\\nonumber
\end{eqnarray}
where $M_i$, $b_i$  and $\omega_{r,i}$ are the total mass, relative radius 
(radius divided by semimajor axis a) and rotational angular 
velocity of component $i$, respectively. $\omega_{k}$ is orbital angular velocity and $e$ is
eccentricity. Functions $g(e)$ and  $f_2(e)$
are given below:
\begin{eqnarray}
g(e)&=&\frac{1}{(1-e^2)^2},\\
f_2(e)&=&\frac{1}{(1-e^2)^5}\left(1+\frac{3}{2}e^2+\frac{1}{8}e^4\right), \nonumber\\\nonumber
\end{eqnarray}
The ratio $\omega_{r,i}/\omega_{k}$ occurring in equation (A.2) 
are given by 
Kopal (1978) as
\begin{equation}
\left(\frac{\omega_{r,1}}{\omega_{k}}\right)^2=\left(\frac{\omega_{r,2}}{\omega_{k}}\right)^2=\frac{1+e}{(1-e)^3}.
\end{equation}

In addition to classical term, secular advance of apsides arises also from the 
general relativistic framework (Kopal 1978)
\begin{equation}
\dot{\omega}_{rel}=6.35\times~10^{-6} \frac{M_1+M_2}{a(1-e^2)}
\end{equation}
where the masses of the component stars ($M_{1}$, $M_{2}$)  are expressed in solar
units and unit of $\dot{\omega}_{rel}$ is $^o/cyc$. 
Then, the total apsidal advance is the summation of $\dot{\omega}_{rel}$ and 
$\dot{\omega}_{cl}$.
}

\subsection{Uncertainty in $\dot{\omega}_{cl}$}

Typical uncertainty in $\dot{\omega}_{cl}$ ($\Delta \dot{\omega}_{cl}$) is computed in terms of uncertainties in masses ($\Delta M_{\rm i}$) and radii ($\Delta R_{\rm i}$) of component stars.
We take care the first row of equation (A.2) and multiply it by 2. Then,
\begin{equation}
\frac{\Delta \dot{\omega}_{cl}}{\dot{\omega}_{cl}}=2\left(\frac{\Delta k_{21}}{k_{21}}+5\frac{\Delta r_{1}}{r_{1}}+\frac{\Delta s}{s}\right)
\end{equation}
where $s$ is the term in the square brackets in th first line of equation (A.2). 
We find $\Delta s$ as
\begin{equation}
\Delta s = \left(15 f_2(e)+\left(\frac{\omega_{r,1}} {\omega_{k}}\right)^2 g(e) \right)
\left( \frac{\Delta M_{2}}{M_{1}+\Delta M_{1}M_2}{M_{1}^2}\right).
\end{equation}
Uncertainty in $k_{21}$ is computed by using
\begin{equation}
\frac{\Delta k_{21}}{k_{21}}=\frac{\partial \log k_{21}}{\partial \log M_{1}}\frac{\Delta M_{1}}{M_{1}} + \frac{\partial \log k_{21}}{\partial \log R_{1}} \frac{\Delta R_{1}}{R_{1}}.
\end{equation}
The partial derivatives ${\partial \log k_{21}}/{\partial \log M_{1}}$ and 
${\partial \log k_{21}}/{\partial \log R_{1}}$ are derived from the models as 0.52 and 1.20, respectively.

\section[t]{Sample Online Tables}
\begin{table*}
\caption{
Basic properties of models with masses from 0.74 to 10 M$_\odot$. See Section 3.
}
$
\renewcommand{\arraystretch}{0.8}
\begin{array}{ccccccc}
\hline
            \noalign{\smallskip}
\log(t/{\rm yr})&\log(R/{\rm R}_\odot) & \log(L/{\rm L}_\odot) & \log(T_{\rm eff}/{\rm K})&\log(k_2)& X_{\rm c}& M/{\rm M}_\odot\\
\hline
  8.09272& -0.17027& -0.76331&3.65605&-1.42985& 0.70090 & 0.74   \\
  9.15473& -0.16198& -0.74999&3.65523&-1.42599& 0.66790 & 0.74   \\
  9.45561& -0.15901& -0.73637&3.65715&-1.43185& 0.63330 & 0.74   \\
  9.64982& -0.15598& -0.72108&3.65946&-1.43923& 0.59480 & 0.74   \\
  9.77217& -0.15335& -0.70668&3.66175&-1.44649& 0.56030 & 0.74   \\
  9.86747& -0.15064& -0.69184&3.66411&-1.45413& 0.52580 & 0.74   \\
  9.94557& -0.14781& -0.67649&3.66652&-1.46203& 0.49140 & 0.74   \\
 10.01157& -0.14481& -0.66059&3.66900&-1.47020& 0.45710 & 0.74   \\
 10.05843& -0.14228& -0.64740&3.67103&-1.47709& 0.42980 & 0.74   \\
 10.10037& -0.13961& -0.63377&3.67311&-1.48408& 0.40260 & 0.74   \\
           \noalign{\smallskip}
            \hline
\end{array}
\renewcommand{\arraystretch}{1.0}
$
\end{table*}

\begin{table*}
\caption{
Isochrones for the time interval from $\log (t{/\rm yr})=6.90-10.25$. See Section 3.
}
$
\renewcommand{\arraystretch}{0.8}
\begin{array}{ccccccccc}
\hline
            \noalign{\smallskip}
\log(t/{\rm yr}) & \log(R/{\rm R}_\odot) & \log(L/{\rm L}_\odot) & \log(T_{\rm eff}/{\rm K})&   M_V   &   B-V  &  U-B    &   M/{\rm M}_\odot &   X_{\rm c}   \\
\hline
  6.90000& 0.64389& 3.80952 &4.39218&-2.28632&-0.24388&-0.92982 &9.89000 &0.54960   \\
  6.90000& 0.64020& 3.79264 &4.38981&-2.25679&-0.24304&-0.92622 &9.78000 &0.55560   \\
  6.90000& 0.62552& 3.75801 &4.38848&-2.17577&-0.24259&-0.92316 &9.62000 &0.57860   \\
  6.90000& 0.57630& 3.52852 &4.35572&-1.78081&-0.23108&-0.87140 &8.27000 &0.60500   \\
  6.90000& 0.56214& 3.47085 &4.34838&-1.67694&-0.22830&-0.85958 &7.98000 &0.61730   \\
  6.90000& 0.55339& 3.43113 &4.34283&-1.60865&-0.22606&-0.85072 &7.78000 &0.62390   \\
  6.90000& 0.53579& 3.34553 &4.33023&-1.46585&-0.22054&-0.83081 &7.36000 &0.63180   \\
  6.90000& 0.52335& 3.28526 &4.32138&-1.36567&-0.21661&-0.81662 &7.08000 &0.63850   \\
  6.90000& 0.50990& 3.20676 &4.30848&-1.24364&-0.21109&-0.79597 &6.72000 &0.63910   \\
           \noalign{\smallskip}
            \hline
\end{array}
\renewcommand{\arraystretch}{1.0}
$
\end{table*}
\section[t]{Figures}
   \begin{figure}[htb]
\centerline{\psfig{figure=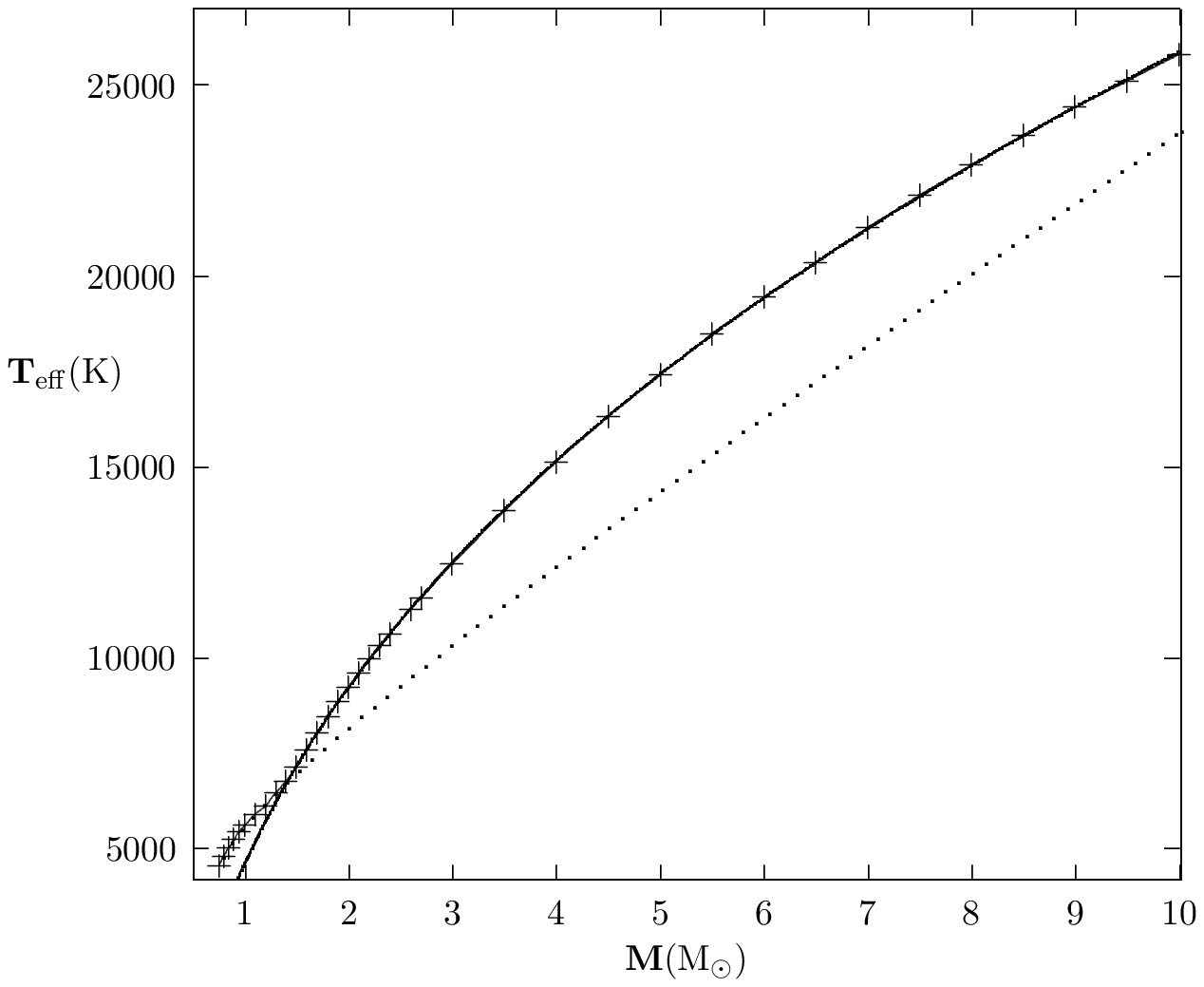,width=200bp,height=230bp}}
\caption{
Effective temperature at ZAMS (+) is plotted with respect to stellar mass. The solid and dotted lines 
show the fitting formula given in equations (6) and (9), respectively. 
}
   \end{figure}
   \begin{figure}[htb]
\centerline{\psfig{figure=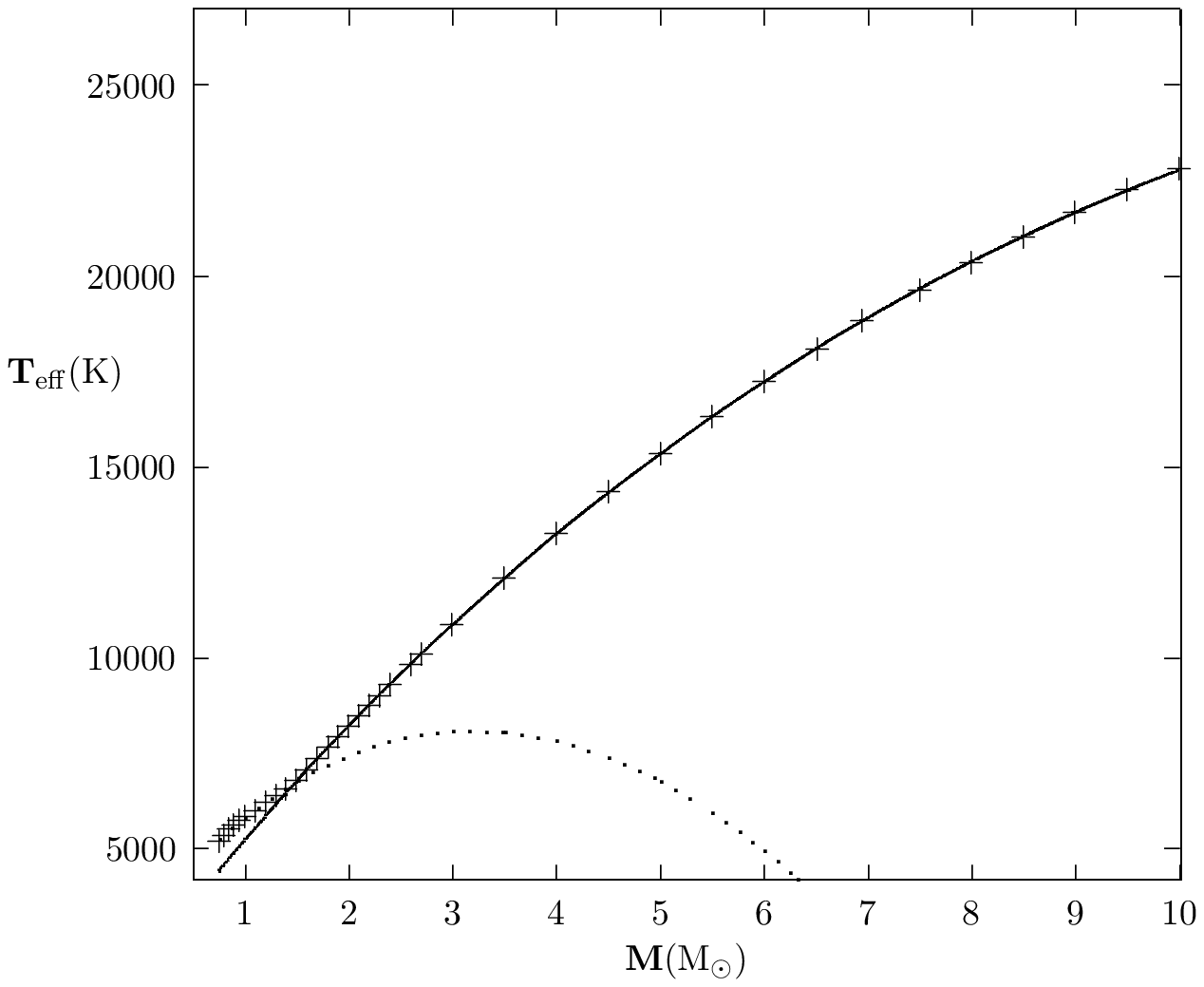,width=200bp,height=230bp}}
\caption{
Effective temperature at TAMS (+) is plotted with respect to stellar mass. The solid and dotted lines 
show the fitting formula given in equations (5) and (10), respectively. 
}
   \end{figure}
   \begin{figure}[htb]
\centerline{\psfig{figure=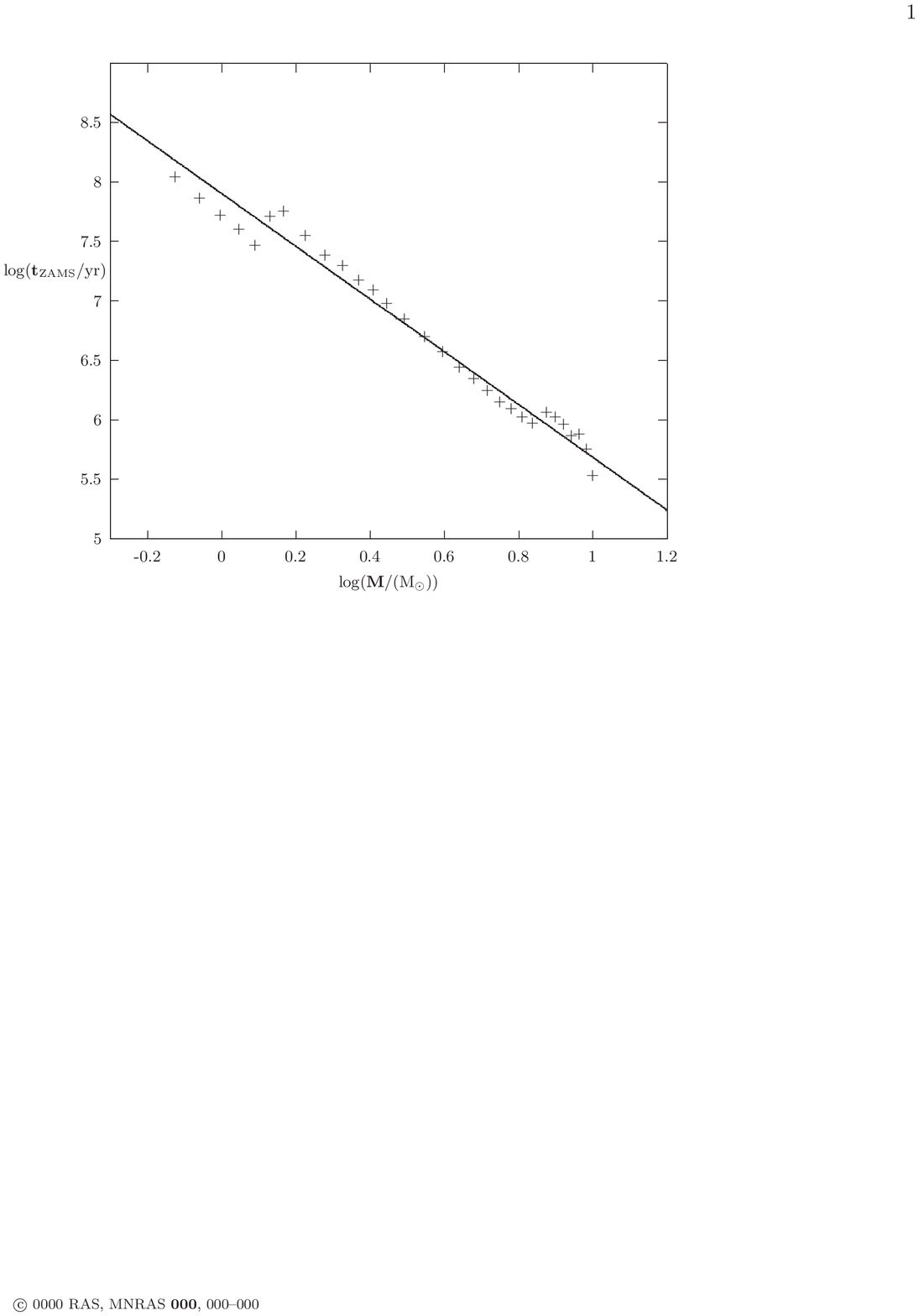,width=200bp,height=230bp}}
\caption{
ZAMS age (+) is plotted with respect to stellar mass. The solid line 
shows the fitting formula given in equations (1). 
}
   \end{figure}

\end{document}